\documentclass[10pt,twocolumn]{article}
\usepackage{flushend,cuted}
\usepackage{geometry}
\usepackage[colorlinks,linkcolor=blue,urlcolor=blue,citecolor=blue]{hyperref}
\usepackage{color}
\usepackage{bm}
\usepackage{amsmath}
\usepackage{graphicx}
\usepackage{float}
\usepackage{subfigure}
\usepackage{epstopdf}
\usepackage{indentfirst}
\usepackage{changepage}
\usepackage{CJK}
\makeatletter
\renewcommand\@seccntformat[1]{{\csname the#1\endcsname}\hspace{0.5em}}
\makeatother
\usepackage{caption2}

\usepackage[numbers,sort&compress]{natbib}

\begin{document}
\title{\Large\textbf{Phase-dependent quantum correlation in cavity-atom system}}
\author{\normalsize{Miaodi Guo, Hongmei Li, Rui Zhang and Xuemei Su}\thanks{Corresponding author: suxm@jlu.edu.cn}\\
\\
\small\emph{Key Lab of Coherent Light, Atomic and Molecular Spectroscopy, Ministry of Education;}\\
\small\emph{College of Physics, Jilin University, Changchun 130012, China}}
\date{}
\maketitle

\begin{strip}
\vspace{-55pt}
\begin{adjustwidth}{1.5cm}{1.5cm}
\begin{abstract}
We propose a scheme to manipulate quantum correlation of output lights from two sides of a cavity by phase control. A probe laser is set to split into two beams in an interferometer with a relative phase in two arms which drive the cavity mode in opposite directions along cavity axis, individually. This phase, here named as driving-field phase, is important to build up quantum correlation in HBT (Hanbury Brown-Twiss) setup. Three control lasers propagate vertically to the cavity axis and drive the corresponding atomic transitions with a closed-loop phase. This type of closed-loop phase has been utilized to realize quantum correlation and even quantum entanglement of the atomic system in previous work [\href{https://doi.org/10.1103/PhysRevA.81.033836}{Phys. Rev. A 81 033836 (2010)}]. The scheme here is useful to manipulate steady and maximum quantum correlation.
\end{abstract}
\vspace{8pt}


\vspace{4pt}
\noindent PACS number(s): 42.50.-p, 42.50.Pq, 32.80.Qk, 42.25.Bs
\end{adjustwidth}
\end{strip}

\section{Introduction}\label{intro}
Electromagnetically induced transparency (EIT), one kind of quantum interference, is based on the coherent superposition of two ground-state levels which are connected with an excited level by two lasers \cite{OL31/2625}. This technique has been realized by increasing the intensity of one of the two lasers as a coupling field ($\Lambda$-type, V-type and ladder-type EIT configurations) \cite{PRA51/576,RMP77/633,PRL100/173602} and has various applications by introducing another level and a third laser as a controlling field \cite{PRL81/3611,PRL97/063901,OL21/1936,OL26/548,PRA68/041801}. Especially, it is becoming more and more important to realize and modify intracavity EIT \cite{OL23/295,PRA82/033808,PRA84/043821,PRA87/053802,PRA89/023806,OL33/46,PRA85/013840,OC358/73} in cavity quantum electrodynamics (cavity-QED) systems \cite{Nature465/755,RMP87/1379}. Comparing with intracavity EIT itself, the modified intracavity EIT has more advantages on light controlling light system, such as low light intensity of switching field and high switching efficiency \cite{PRA85/013814,PRA85/013840,OC358/73,Annp1700427}.

On the other hand, there are a lot of interesting reports on phase-dependent EIT induced by a closed loop \cite{PRA59/2302,PLA324/388,PRA71/011803,MOP54/2459} with or without cavity, which turn out to be a more sophisticated manner in manipulating light-atom interactions. It is applied in many explorations, such as phase-control spontaneous emission \cite{PRL81/293}, beam splitter \cite{PRL101/043601} and entanglement between collective fields \cite{PRA81/033836}. Based on phase-dependent EIT in a double four-level-atom system \cite{PRA81/033836}, four incident fields act as two collective modes and constitute two quantum beats. As a result, entanglement occurs between two collective modes. One can also operate the nonadiabatic optical transitions, quantum mechanical superposition states, the polarization selection and even controllable phase gate \cite{PRA79/025401,PRA87/013430,PRA90/063841,PRA92/043838} by phase control in closed-loop EIT configuration. Meanwhile, with another type of phase control in the two arms of  quantum interferometer, a lot of creative applications can be realized. For an example, with a Kerr phase in one arm of quantum interferometer, the postselected measurement and the amplification of weak effect have been carried out where giant XPM (cross-phase modulation) nonlinearity is resulted from EIT technique \cite{PRA90/013827}. The phase difference existing in two arms of quantum interferometer takes important effect on sensitive measurement like group-delay measurement and gravitational waves detection \cite{PRA85/011801,OL20/788,PREP684/1}. Moreover, this kind of phase difference is involved in interference control of medium absorption, measurement of spatial correlation or entanglement in many literatures \cite{PRL105/053901,Science331/889,PRA92/023824,PRA95/013841}. In the scheme here, we utilize this driving-field phase to manipulate output of cavity mode field at two ends to realize controllable quantum correlation, where dissipation of the system is controlled by the modified EIT  with closed-loop phase. Here controlled quantum coherence is induced by phase coherence of the classical fields.

In the following, we firstly make an introduction to our scheme. Some four-level atoms trapped in the cavity interact with three control lasers inducing a closed-loop phase and a probe laser splits into two coherent beams which drive the cavity mode field with a driving-field phase. Then we make a theoretical analysis and obtain the analytic solutions for intracavity field and its output fields. We can make the incident probe laser perfectly absorbed in cavity or totally transmitted (reflected) out of the cavity. And the correlation between two output channels depends on the closed-loop and driving-field phases. Secondly, we move forward on the numerical results and detailed discussion in section 3. We analyze how our system is performed as a perfect photon absorber or a complete transmitter / reflector. We discuss the controlled effects of the relative phases on quantum correlation according to evolution of second-order correlation of two output channels. In the end, we make a simple summary in section 4. 

\section{Theoretical analysis}\label{theory}
\begin{figure}[!hbt]
\centering
\subfigure{
\begin{minipage}{6.5cm}
\includegraphics[width=6cm,bb=48 385 470 770]{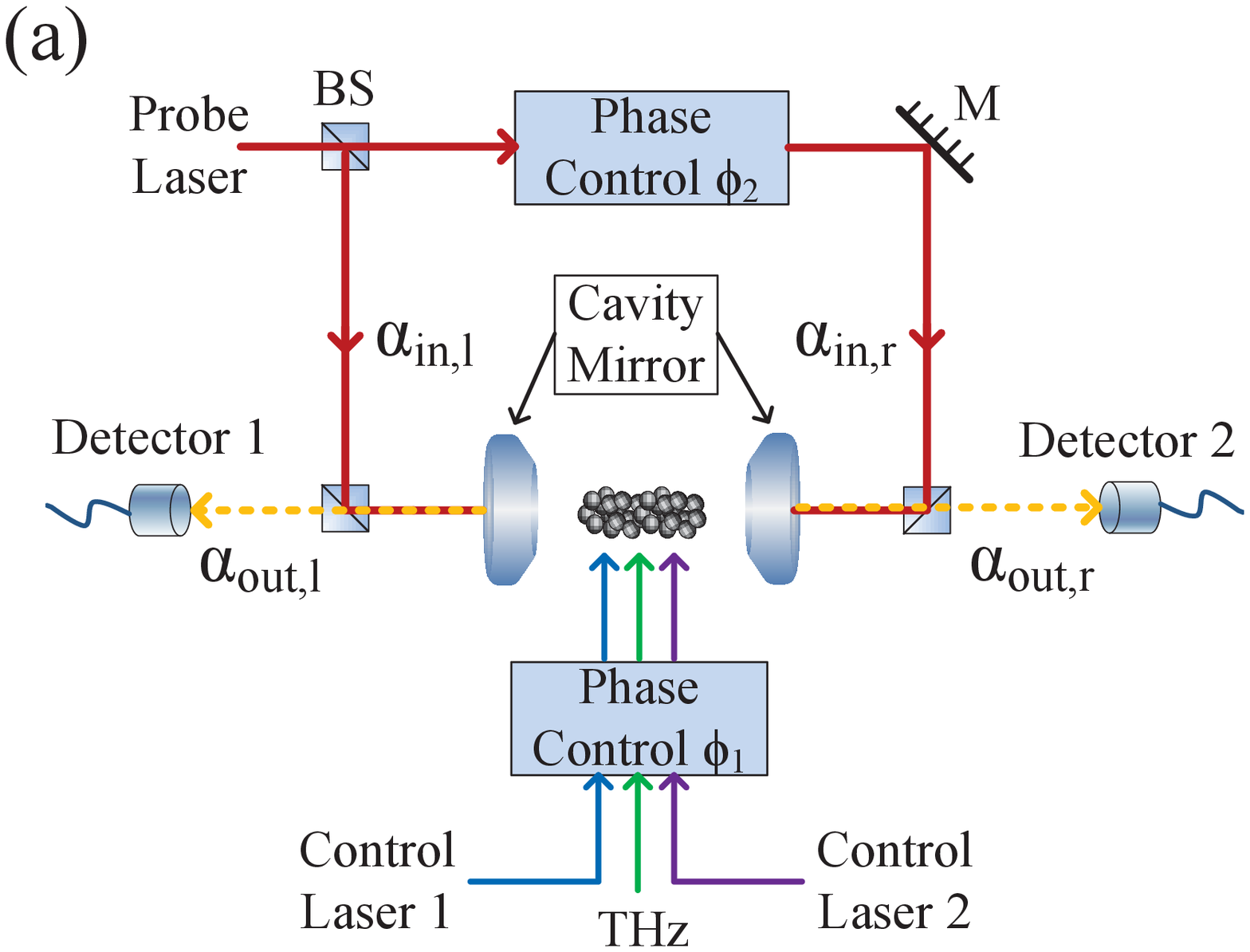}
\end{minipage}}
\subfigure{
\begin{minipage}{7cm}
\includegraphics[width=6cm,bb=50 470 470 770]{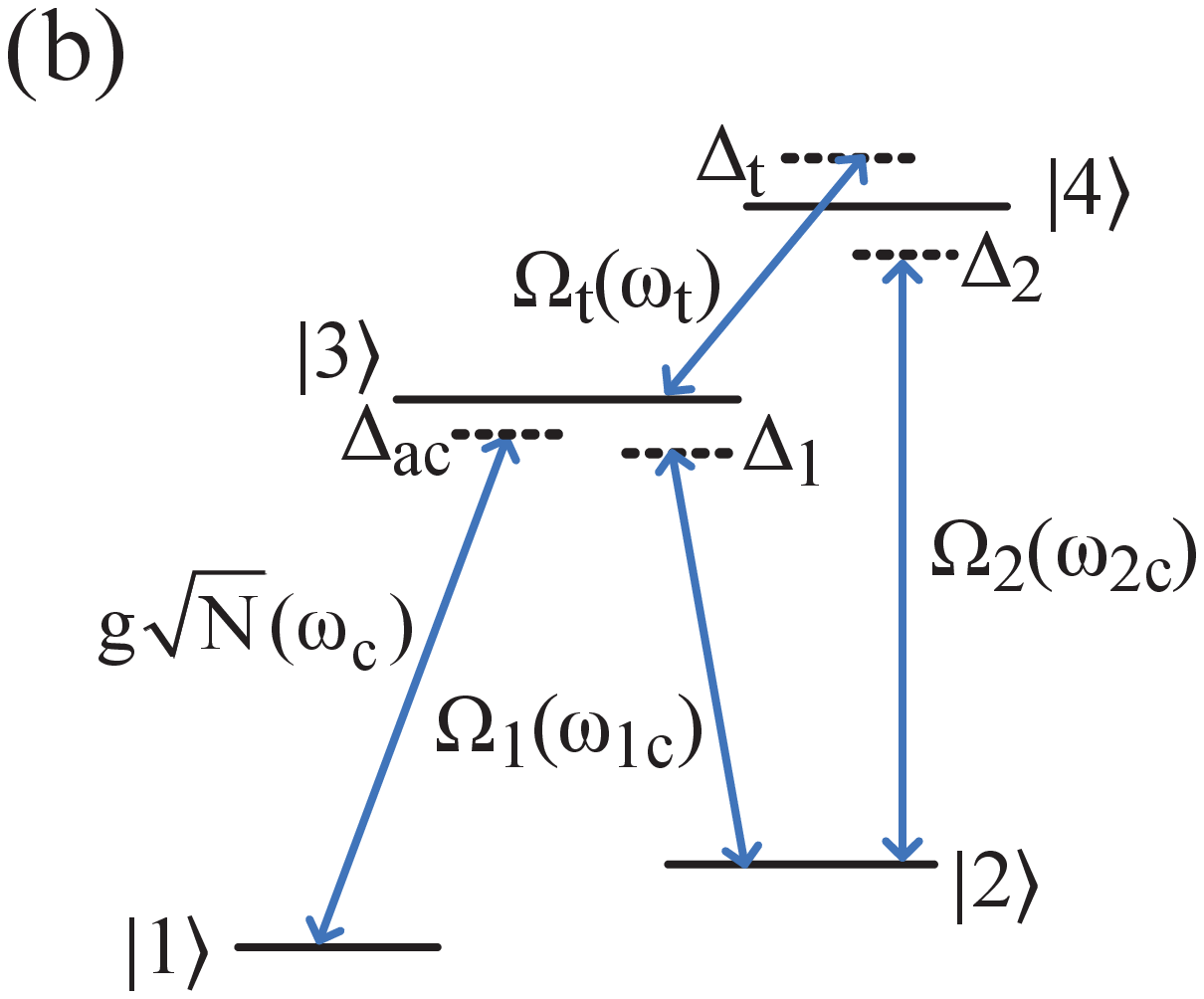}
\end{minipage}}
\vspace{-5pt}
\caption{(a) The theoretical scheme and (b) the four-level atomic level structure of $^{87}$Rb. }\label{fig-system}
\end{figure}
The scheme proposed here is depicted as in Fig. \ref{fig-system}(a). The atomic levels $|1\rangle$, $|2\rangle$, $|3\rangle$ and $|4\rangle$ as shown in figure \ref{fig-system}(b) correspond to $5S_{1/2}\;F=1$, $5S_{1/2}\;F=2$, $5P_{1/2}\;F=1$ and $5P_{3/2}\;F=2$ of $^{87}$Rb, respectively. Two control lasers and a terahertz (THz) wave enter into the cavity with closed-loop phase $\phi_{1}$. Those two control lasers drive the atomic transitions $|2\rangle\rightarrow|3\rangle$, $|2\rangle\rightarrow|4\rangle$ with frequency detuning $\Delta_{1}=\omega_{1c}-\omega_{32}$ and $\Delta_{2}=\omega_{2c}-\omega_{42}$, respectively. The THz wave as a third control laser couples atomic transition from level $|3\rangle$ to $|4\rangle$ with a frequency detuning $\Delta_{t}=\omega_{t}-\omega_{43}$. A probe laser ($\omega_{p}$), which has a frequency detuning $\Delta_{c}=\omega_{p}-\omega_{c}$ from cavity mode ($\omega_{c}$), is split by a beam splitter (BS). The two split beams $\alpha_{in,l}$ and $\alpha_{in,r}$ are injected into cavity from the opposite directions. With phase control device, a relative phase $\phi_{2}$, namely driving-field phase, exists between $\alpha_{in,l}$ and $\alpha_{in,r}$.  Two detectors are applied to receive output signal from right and left cavity mirror. $\Delta_{ac}=\omega_{c}-\omega_{31}$ is frequency detuning of cavity mode and atomic transition $|1\rangle\rightarrow|3\rangle$. $g\sqrt{N}$ is the collective coupling coefficient of cavity-QED system. $\Omega_{1}(\omega_{1c})$, $\Omega_{2}(\omega_{2c})$ and $\Omega_{t}(\omega_{t})$ are Rabi frequency (angular frequency) of control laser 1, control laser 2 and THz wave. $\omega_{31}$, $\omega_{32}$, $\omega_{42}$ and $\omega_{43}$ are angular frequency of corresponding atomic level spacing. It is well known that EIT or modified EIT can be observed in this kind of configuration \cite{PLA324/388,CPB26/074207}. The optical switch based on intensity modulation of control lasers can be realized. Different from that, however, here we utilize the phase effect to manipulate the intracavity field and output signals.

Under rotating wave approximation, the Hamiltonian is as following,
\begin{equation}
\begin{split}
H\!\!=\!&\!-\!\hbar\Delta_{c}a^{\dag}a\!-\!\hbar\sum_{j=1}^{N}[(\Delta_{p}\!\!-\!\!\Delta_{1})\sigma_{22}^{j}\!+\!(\Delta_{p}\!\!-\!\!\Delta_{1}\!\!+\!\!\Delta_{2}\!\!-\!\!\Delta_{t})\sigma_{33}^{j}\\
&\!+\!(\Delta_{p}\!\!-\!\!\Delta_{1}\!\!+\!\!\Delta_{2})\sigma_{44}^{j}]\!-\!\hbar\sum_{j=1}^{N}(ga^{\dag}\sigma_{13}^{j}e^{i\varphi_{p}}\!+\!\Omega_{1}\sigma_{23}^{j}e^{i\varphi_{1}}\\
&\!+\!\Omega_{2}\sigma_{24}^{j}e^{i\varphi_{2}}\!+\!\Omega_{t}\sigma_{34}^{j}e^{i\varphi_{t}})\!+\!H.C.,\label{Hamiltonian}
\end{split}
\end{equation}
where $H.C.$ denotes the Hermitian conjugate, $\Delta_{p}=\omega_{p}-\omega_{31}$ is frequency detuning of probe laser and atomic transition $|1\rangle\rightarrow|3\rangle$, $g=\mu_{13}\sqrt{\omega_{c}/(2\hbar\varepsilon_{0}V)}$ is cavity-QED coupling coefficient, $a^{\dag}$($a$) is creation (annihilation) operator of cavity photons, $\sigma_{mn}^{j}=|m\rangle\langle n|$ ($m,n=1,2,3,4$) is atomic operator, $\varphi_{p}$, $\varphi_{1}$, $\varphi_{2}$ and $\varphi_{t}$ are phases of probe laser, control laser 1, control laser 2 and THz wave, respectively.

For simplicity, we consider a symmetric Fabry-Perot cavity with field loss rate $\kappa_{l}$ ($\kappa_{r}$) from left (right) cavity mirror, $\kappa_{i}=T_{i}/2\tau$, where $T_{i}$ is the mirror transmission and $\tau$ is the photon round-trip time inside the cavity. The relation between input and output modes of this cavity-atom system is given by \cite{G.S.Agarwal2013},
\begin{gather}
\begin{split}
&\langle a_{out,l}\rangle+\langle a_{in,l}\rangle=\sqrt{2\kappa_{l}\tau}\;\langle a\rangle,\\
&\langle a_{out,r}\rangle+\langle a_{in,r}\rangle=\sqrt{2\kappa_{r}\tau}\;\langle a\rangle,\label{input-output relation}
\end{split}
\end{gather}
where $\langle{a}\rangle=\alpha$ ($\langle{a^{\dag}}\rangle=\alpha^{*}$), $\langle a_{in,l}\rangle=\alpha_{in}^{l}$ ($\langle a_{in,r}\rangle=\alpha_{in}^{r}$) and $\langle a_{out,l}\rangle=\alpha_{out}^{l}$ ($\langle a_{out,r}\rangle=\alpha_{out}^{r}$) are expectation values for the operators of intracavity field, incident probe beam and outgoing signal from left (right) mirror, respectively \cite{PRA93/023806}. Since the system is driven from both sides, the transmission and reflection properties of the cavity can be calculated by solving the following Heisenberg-Langevin equations of motion \cite{RMP87/1379},
\begin{equation}
\begin{split}
&\langle\dot a\rangle\!\!=\!\!\frac{1}{i\,\hbar}[a,H]\!\!-\!\!(\kappa_{l}\!\!+\!\!\kappa_{r})\langle a\rangle\!\!+\!\!\sqrt{2\kappa_{l}/\tau}\langle a_{in}^{l}\rangle\!\!+\!\!\sqrt{2\kappa_{r}/\tau}\langle a_{in}^{r}\rangle,\\
&\langle\dot \sigma_{ij}\rangle\!\!=\!\!\frac{1}{i\,\hbar}[\sigma_{ij},H]\!\!-\!\!\gamma_{ij}\langle\sigma_{ij}\rangle.\label{equation of motion}
\end{split}
\end{equation}
At the initial time, we assume that the populations in level $|1\rangle$ and $|2\rangle$ are both 1/2, namely $\sigma_{11}=\sigma_{22}=1/2$ and $\sigma_{33}=\sigma_{44}=0$. Then under steady-state condition, the intracavity field can be derived as,
\begin{equation}
\alpha=\frac{\sqrt{2\kappa_{l}/\tau}\;\alpha_{in}^{l}+\sqrt{2\kappa_{r}/\tau}\;\alpha_{in}^{r}}{(\kappa_{l}+\kappa_{r})-i\Delta_{c}-i\chi},\label{intracavity field}
\end{equation}
where
\begin{equation*}
\chi\!=\!\frac{g^{2}N/2(\Omega_{2}^{2}\!-\!A\!*\!B)}{2\Omega_{1}\Omega_{2}\Omega_{t}\cos{\phi_{1}}\!-\!A\!*\!\Omega_{t}^{2}\!-\!B\!*\!\Omega_{1}^{2}\!-\!C\!*\!\Omega_{2}^{2}\!+\!A\!*\!B\!*\!C}
\end{equation*}
is the susceptibility of atomic media and $A=\Delta_{p}-\Delta_{1}+i\,\gamma_{12}$, $B=(\Delta_{p}-\Delta_{1}+\Delta_{2})+i\,\Gamma_{4}/2$, $C=(\Delta_{p}-\Delta_{1}+\Delta_{2}-\Delta_{t})+i\,\Gamma_{3}/2$. Here $\phi_{1}=\varphi_{1}-\varphi_{2}+\varphi_{t}$ represents the relative phase of two control lasers and THz wave induced by the closed loop as shown in Fig. \ref{fig-system}(b), $\Gamma_{3}=\Gamma_{4}=\Gamma$ is the natural decay rate of excited states $|3\rangle$ and $|4\rangle$, and $\gamma_{12}$, much smaller than $\Gamma_{3}$ or $\Gamma_{4}$, is the decoherence rate between ground states $|1\rangle$ and $|2\rangle$.

The analytical solutions to intracavity field and output signals through right and left mirrors are,
\begin{equation}
\begin{split}
&I_{c}=I_{in}^{r}|\frac{\sqrt{\kappa}(1+e^{i\phi_{2}})}{\kappa-i\Delta_{c}-i\chi}|^{2},\\
&I_{out}^{r}=I_{in}^{r}|\frac{\kappa(1+e^{i\phi_{2}})}{\kappa-i\Delta_{c}-i\chi}-1|^{2},\\
&I_{out}^{l}=I_{in}^{r}|\frac{\kappa(1+e^{-i\phi_{2}})}{\kappa-i\Delta_{c}-i\chi}-1|^{2}.\label{field intensity}
\end{split}
\end{equation}
Here we assume that $\kappa_{l}=\kappa_{r}=\kappa/2$, $\alpha_{in}^{l}=|\alpha_{in}|e^{i\varphi_{l}}$ and $\alpha_{in}^{r}=|\alpha_{in}|e^{i\varphi_{r}}$. The driving-field phase $\phi_{2}=\varphi_{l}-\varphi_{r}$ is the relative phase of two incident probe beams, $I_{in}^{r}$ is the input field intensity from right side of cavity and $I_{c}$, $I_{out}^{r}$ ($I_{out}^{l}$) are field intensity of intracavity light and output light from right (left) mirror, respectively. The output field intensity can be manipulated either by changing media susceptibility or relative phase $\phi_{2}$. In this system, both strength / frequency of control lasers and closed-loop phase $\phi_{1}$ can be used to modulate media absorption property. As indicated in equation (\ref{field intensity}) two output channels have correlation based on $\phi_{2}$, especially, when $\phi_{2}=\pi$, intracavity field intensity is always equal to zero, which means the probe beams are transmitted or reflected totally. In brief, in this scheme, light output of the two channels can be controlled by two types of phases.

\section{Results and discussion}\label{result}
In this section, we firstly analyze the media absorption property controlled by intracavity phase-dependent EIT induced by closed-loop phase $\phi_{1}$ since it is one of the major factors of phase-dependent correlation between two output channels. Here we emphasize the effect of initial phases of controlling lasers rather than the intensities of them. Secondly, we discuss influence by the other major factor, driving-field phase $\phi_{2}$. The two phases lead to the controllable second-order correlation and the steady and maximum value of it. Numerical results and detailed discussion are as following.

\begin{figure*}[!hbt]
\centering
\subfigure{
\begin{minipage}{4.2cm}
\includegraphics[width=4.4cm,bb=0 0 555 450]{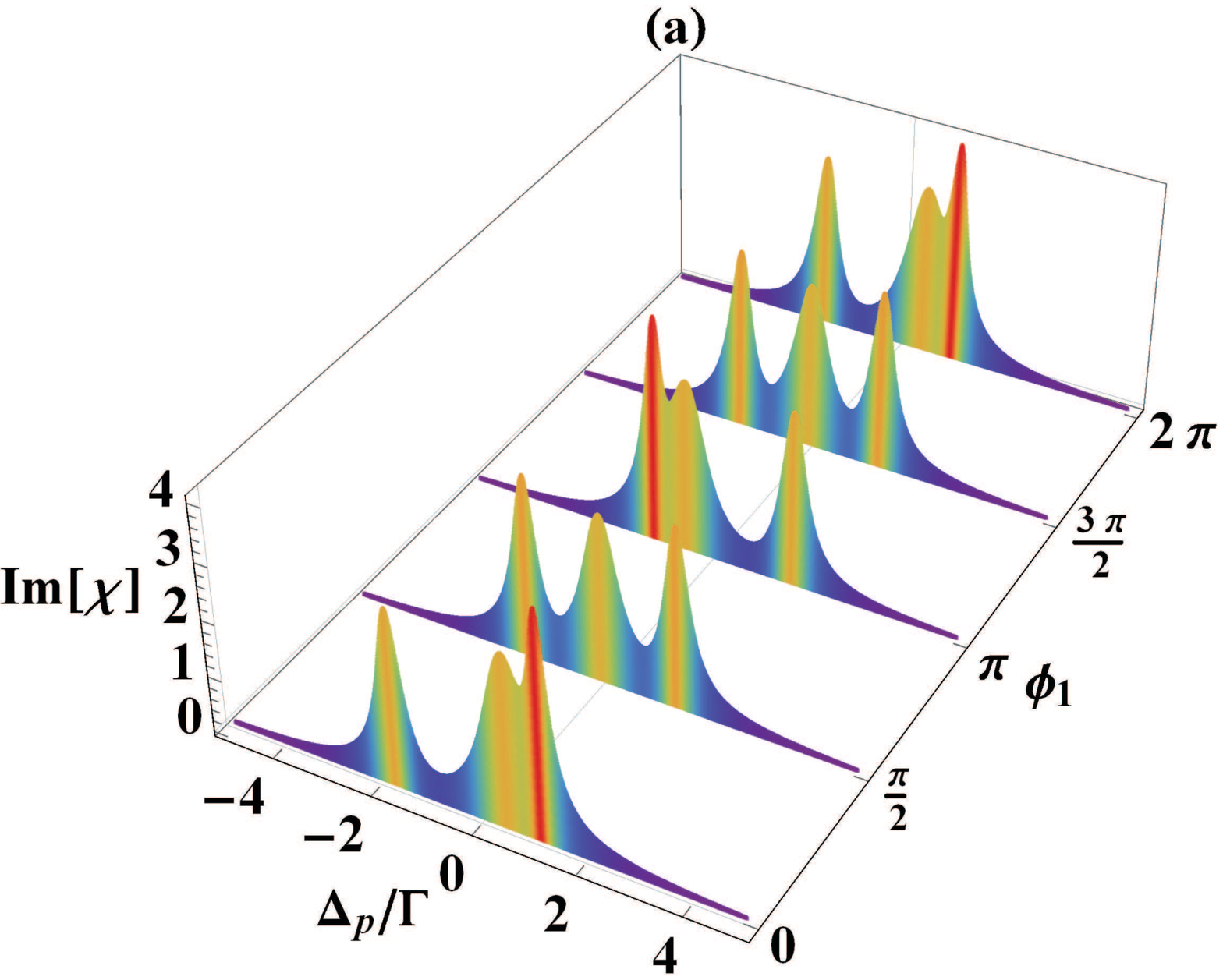}
\end{minipage}}
\subfigure{
\begin{minipage}{4.2cm}
\includegraphics[width=4.4cm,bb=0 0 555 450]{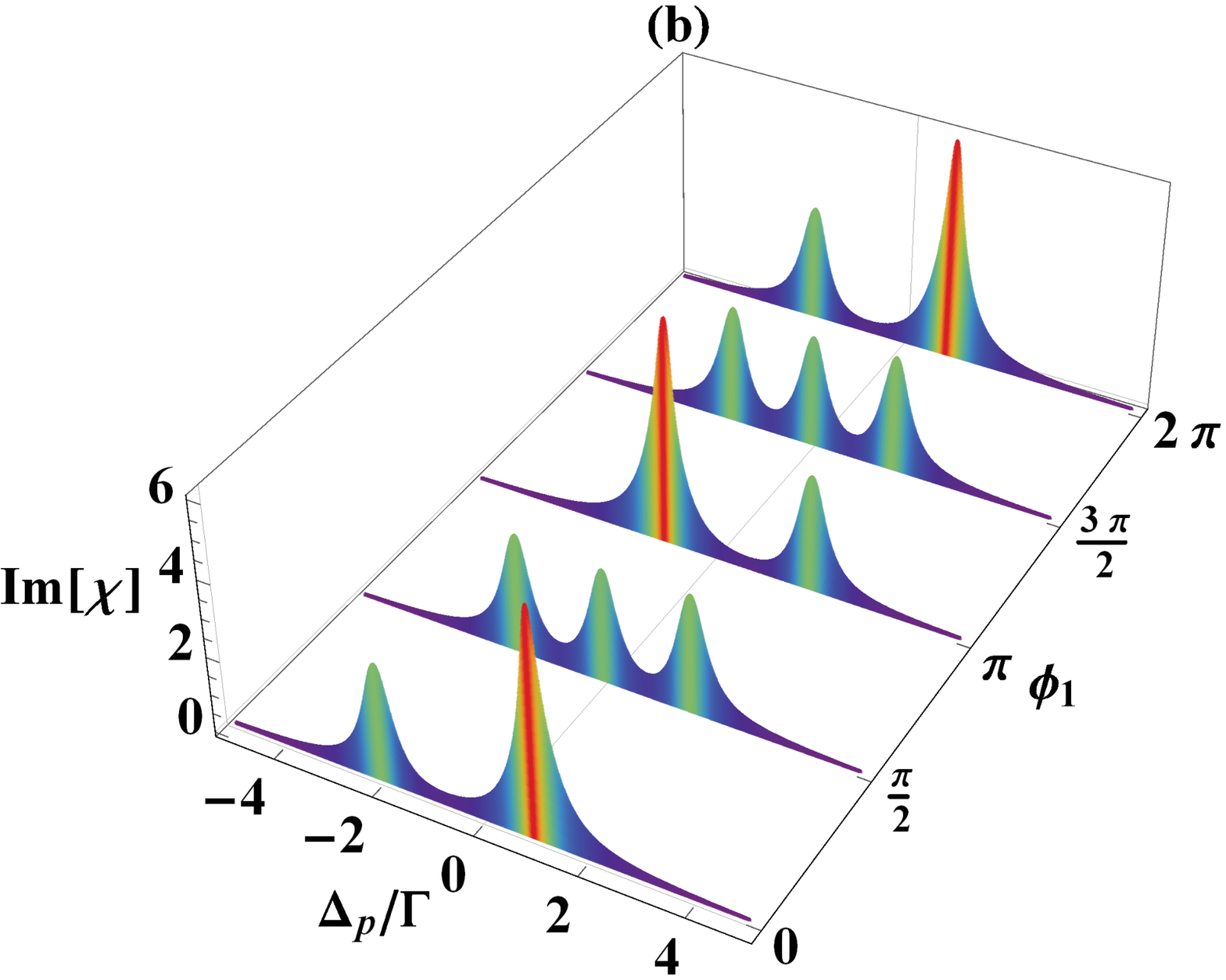}
\end{minipage}}
\subfigure{
\begin{minipage}{4.2cm}
\includegraphics[width=4.4cm,bb=0 0 605 450]{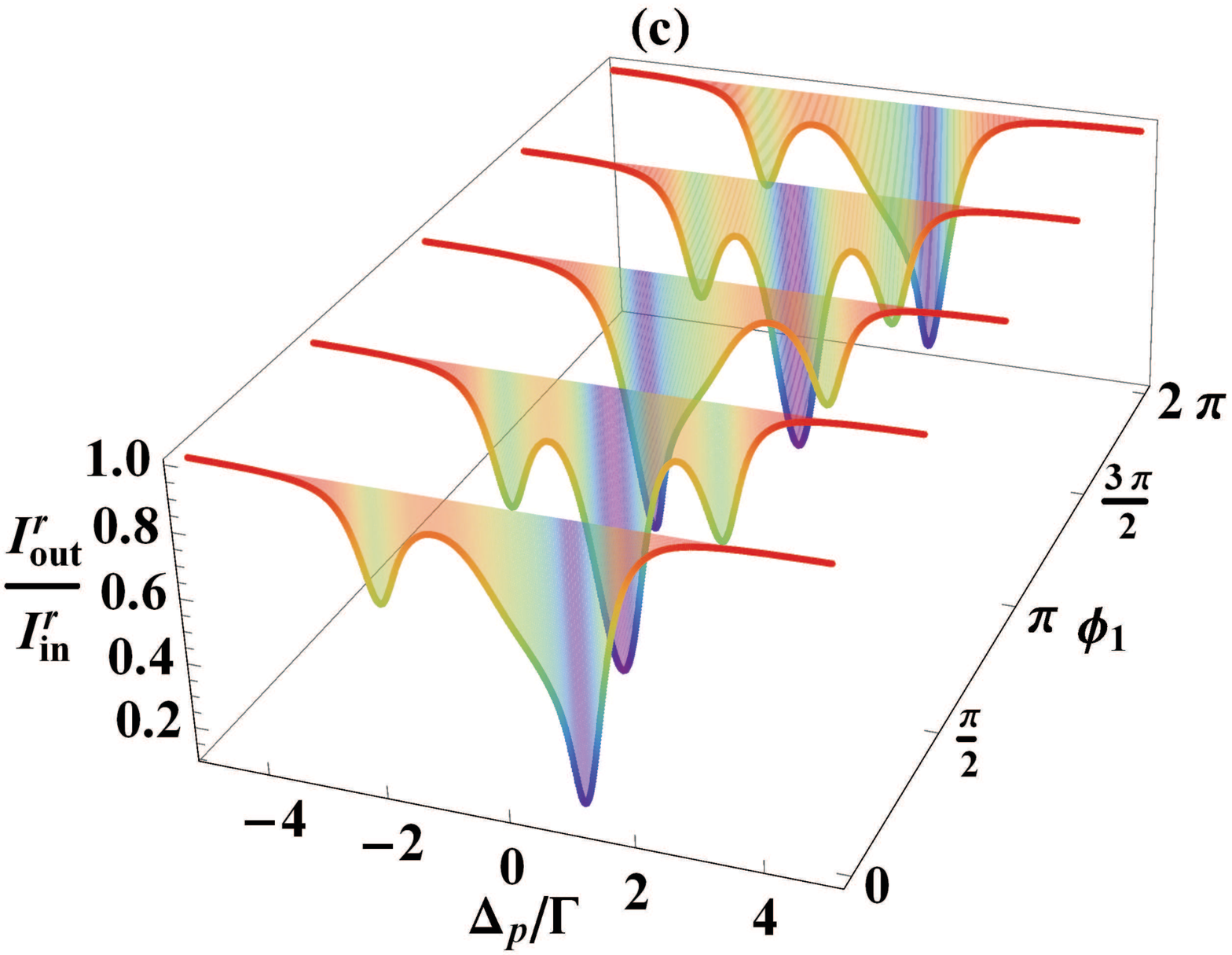}
\end{minipage}}
\subfigure{
\begin{minipage}{4.1cm}
\includegraphics[width=4.4cm,bb=0 0 605 495]{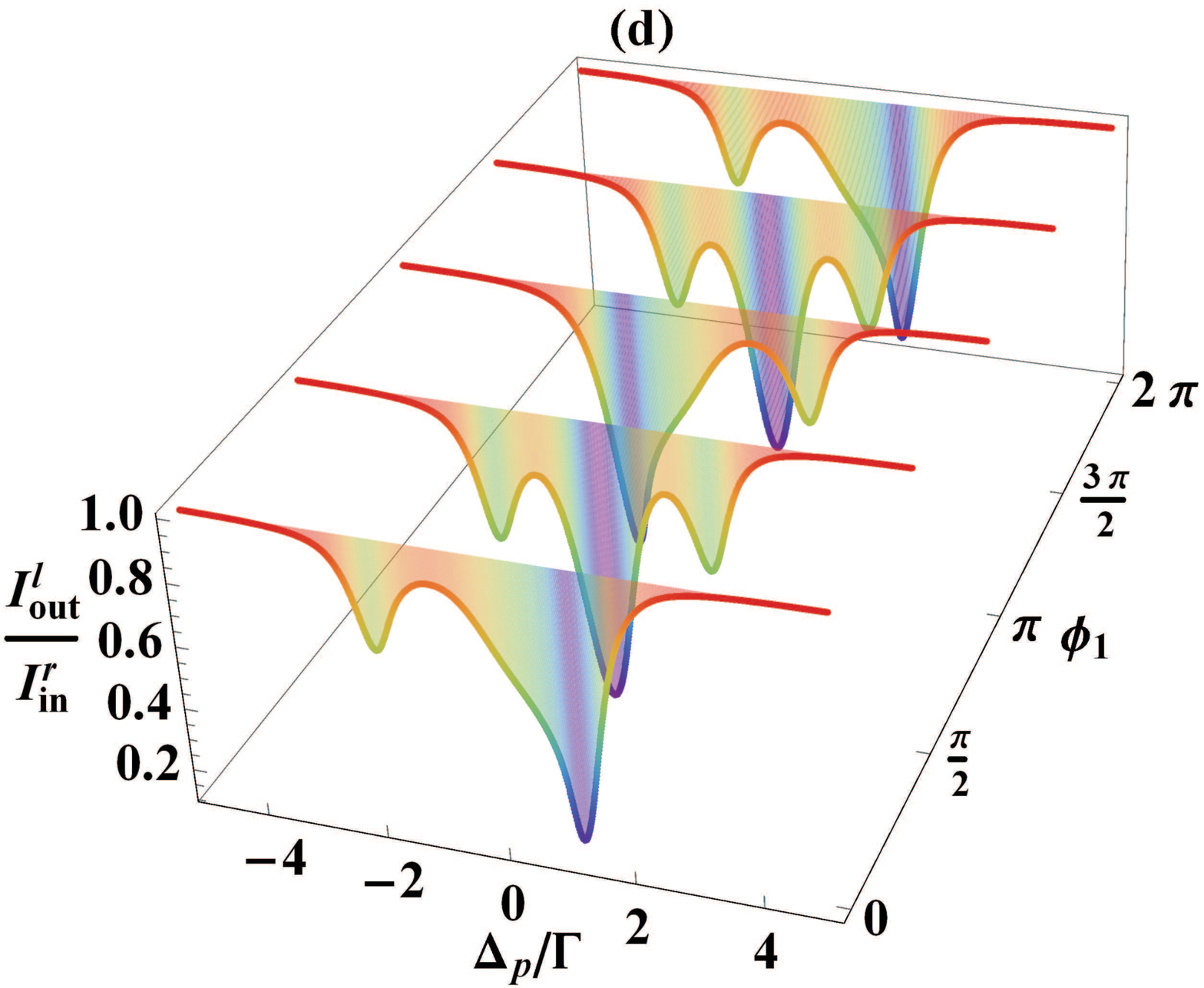}
\end{minipage}}
\vspace{-5pt}
\caption{(a), (b) Absorption property of atomic media inside the cavity and (c), (d) output intensity from two sides of cavity. The parameters are $g\sqrt{N}=\Gamma$, $\Omega_{1}=\Omega_{2}=\Gamma$, $\kappa=\Gamma$, $\gamma_{12}=0.001\Gamma$, and (a) $\Omega_{t}=0.5\Gamma$, (b), (c), (d) $\Omega_{t}=\Gamma$.}\label{fig-imx}
\end{figure*}

Media absorption property is presented by imaginary part of susceptibility $\chi$ as shown in Fig. \ref{fig-imx}(a) and \ref{fig-imx}(b). The parameters of Fig. \ref{fig-imx} and all other figures in this section are under resonance condition $\Delta_{1}=\Delta_{2}=\Delta_{t}=\Delta_{ac}=0$. The cavity is at threshold of strong collective-coupling regime ($g^{2}N=\kappa\,\Gamma$). Without control laser 2 and THz wave, $\Lambda$-type three-level atoms interact with cavity field and control laser 1, and thus two bright polaritons are formed at $\Delta_{p}=\pm\sqrt{g^{2}N+\Omega_{1}^{2}}$ and a dark state at $\Delta_{p}=0$ which is decoupled to cavity mode. When two control lasers are applied, two new dark states because of dark state splitting and new bright states are formed at \cite{PRA85/013814}
\begin{equation*}
\begin{split}
&\Delta_{p}=\\
&\pm\sqrt{[\Omega_{1}^2\!+\!\Omega_{2}^2\!+\!g^2N\pm\sqrt{(\Omega_{1}^2\!+\!\Omega_{2}^2\!+\!g^2N)^2\!-\!4g^2N\Omega_{2}^2}]/2}.
\end{split}
\end{equation*}
Together with the strong THz wave ($\Omega_{t}$), intracavity EIT splitting will be destroyed but a closed interaction contour is formed which can lead to a phase-dependent EIT.
\begin{figure*}[!hbt]
\centering
\subfigure{
\begin{minipage}{5.5cm}
\includegraphics[width=5cm]{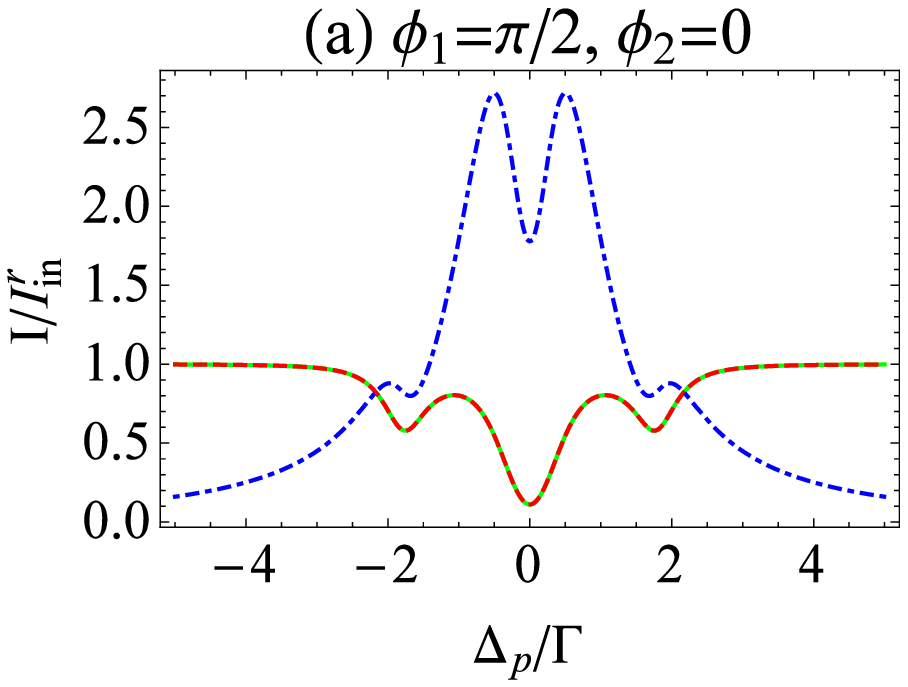}
\end{minipage}}
\subfigure{
\begin{minipage}{5.5cm}
\includegraphics[width=5cm]{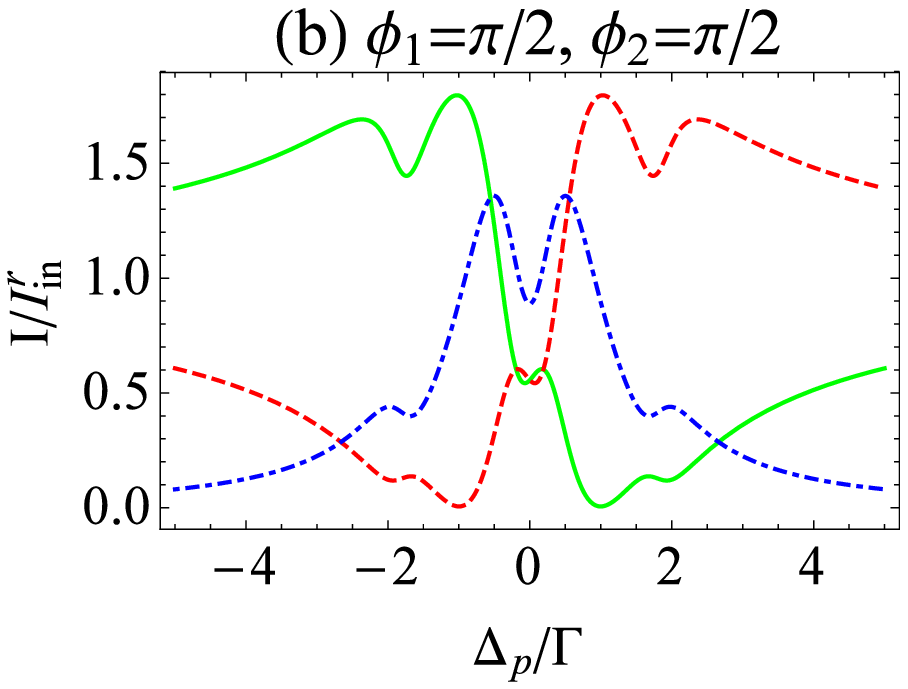}
\end{minipage}}
\subfigure{
\begin{minipage}{5.5cm}
\includegraphics[width=5cm,bb=0 -5 260 200]{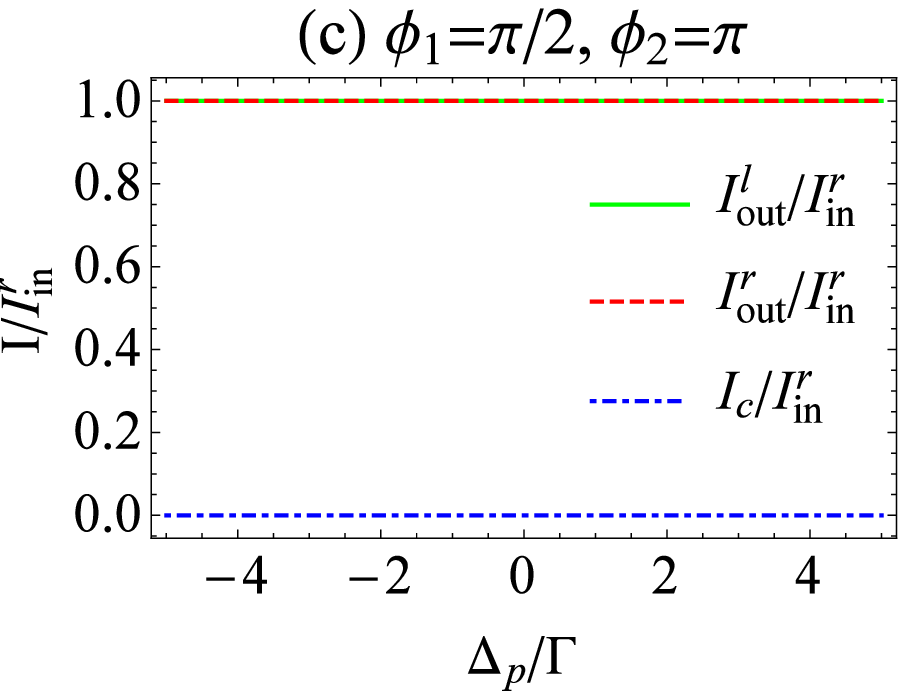}
\end{minipage}}
\vspace{-10pt}
\caption{Field intensity ratio versus frequency detuning $\Delta_{p}$ for output field from left ($I_{out}^{l}/I_{in}$ (green solid lines)) or right mirror ($I_{out}^{r}/I_{in}$ (dashed red lines)) and intracavity field ($I_{c}/I_{in}$ (dot-dashed blue lines)). The parameters are $\phi_{1}=\pi/2$ for (a) $\phi_{2}=0$, (b) $\phi_{2}=\pi/2$, and (c) $\phi_{2}=\pi$. The other parameters are the same as in Fig. \ref{fig-imx}(b). }\label{fig-intensity}
\end{figure*}

Fig. \ref{fig-imx}(a) and \ref{fig-imx}(b) are plotted with $\Omega_{t}=0.5\Gamma$ and $\Omega_{t}=\Gamma$ to present the harmful effect of THz wave on intracavity EIT splitting and the compensated contribution of the closed-loop phase. In figure \ref{fig-imx}(a) when $\Omega_{t}=0.5\Gamma$, THz wave could not completely destroy intracavity EIT splitting. This is why Im[$\chi$] has three peaks in absorption spectra (double EIT), but the values and locations of the three peaks are decided by closed-loop phase. As shown in Fig. \ref{fig-imx}(a), for $\phi_{1}$ in even order of $\pi/2$ ($\phi_{1}=0,\,\pi,\,2\pi$) and odd order of $\pi/2$ ($\phi_{1}=\pi/2,\,3\pi/2$), there are three asymmetrical and symmetrical absorption peaks with respect to $\Delta_{p}=0$, respectively. The basic principle behind this phenomenon can be explained from the expression of Im[$\chi$] as below,
\begin{equation}
Im[\chi]\!=\!g^2N
\begin{cases}
\frac{X-8(\Gamma+2\gamma_{12})\Omega_{1}\Omega_{2}\Omega_{t}\Delta_{p}}{[Z_{1}-4(\Gamma+\gamma_{12})\Delta_{p}^2]^2+Y_{1}^2},&{\phi_{1}\!=\!0,\pi,2\pi}\\
\\
\frac{X}{[Z_{1}-4(\Gamma+\gamma_{12})\Delta_{p}^2]^2+Y_{2}^2},&{\phi_{1}\!=\!\pi/2,3\pi/2}
\end{cases}\label{imx}
\end{equation}
where
\begin{equation*}
\begin{split}
&X=4\Gamma\Delta_{p}^4+Z_{1}(\Gamma\gamma_{12}+2\Omega_{2}^2)+Z_{2}\Delta_{p}^2,\\
&Y_{1}=Y_{2}+8\Omega_{1}\Omega_{2}\Omega_{t},\\
&Y_{2}=4\Delta_{p}^3-\Delta_{p}[\Gamma^2+4\Gamma\gamma_{12}+4(\Omega{1}^2+\Omega_{2}^2+\Omega_{t}^2)],\\
&Z_{1}=\Gamma^2\gamma_{12}+2\Gamma(\Omega_{1}^2+\Omega_{2}^2)+4\gamma_{12}\Omega_{t}^2,\\
&Z_{2}=\Gamma^3+8\gamma_{12}\Omega_{1}^2+4\Gamma(\gamma_{12}^2-2\Omega_{2}^2+\Omega_{t}^2).
\end{split}
\end{equation*}

When $\Omega_{t}$ is up to $\Gamma$, it stops the dark-state splitting completely for $\phi_{1}$ in the even order of $\pi/2$. While $\phi_{1}$ is equal to odd order of $\pi/2$, three symmetrical absorption peaks can exhibit as shown in Fig. \ref{fig-imx}(b), and thus the double EIT reappears. These phase-dependent EIT properties have also been presented in the output spectra as shown in \ref{fig-imx}(c) and \ref{fig-imx}(d) when $\phi_{2}=0$, which show that dissipation of the system can be modulated by closed-loop phase.

In the following, we explore how $\phi_{2}$ affects output fields from two channels. The output intensities of two sides of the cavity and that of intracavity field are depicted in Fig. \ref{fig-intensity} under several $\phi_{2}$ when double EIT exists ($\phi_{1}=\pi/2$). Fig. \ref{fig-intensity}(a) shows that the output intensity from right side of cavity mirror is always the same as that from left side at any frequency detuning. It shows that, at CPA (coherent perfect absorber) resonance ($\Delta_{p}=0$), large absorption and less interference amplitudes between probe beams lead to an ideal interference ``trap" for the two beams so that eventually the probe photon will be absorbed by intracavity media. With increasing relative phase $\phi_{2}$, due to interference between two probe beams, output of intracavity field is enhanced at resonance (as in Fig. \ref{fig-intensity}(b)) until total reflection ($I_{out}/I_{in}=1$) of the intracavity field appears at two ends of the cavity under $\phi_{2}=\pi$ as shown in \ref{fig-intensity}(c), which can be derived by equation (\ref{field intensity}).

\begin{figure}
\centering
\includegraphics[width=7cm,bb=15 9 800 520]{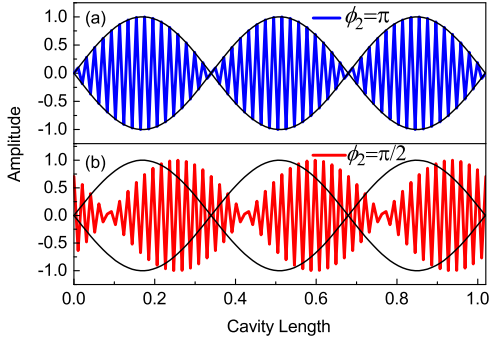}
\caption{Field distribution in two-sided cavity with (a) $\phi_{2}=\pi$ and (b) $\phi_{2}=\pi/2$.}\label{fig-intendis}
\end{figure}

The physical essence of the above effect of $\phi_{2}$ is resulted from formation of a phase-dependent standing wave by two driving fields inside of cavity. When $\phi_{2}=\pi$, the standing wave (blue line in Fig. \ref{fig-intendis}(a)) takes concerted action with the cavity mode field (black lines in Fig. \ref{fig-intendis}), thus the intracavity photon will be totally reflected by standing-wave field. When $\phi_{2}$ is equal to other values, for example $\phi_{2}=\pi/2$, the standing-wave field (red line in Fig. \ref{fig-intendis}(b)) will always be out of step with cavity mode field, therefore intracavity field will be partially reflected through two ends of cavity. A light switching with high contrast can be designed with different driving-field phase and closed-loop phase. With $\phi_{1}=\pi/2$, for example, the two output channels at resonance are closed when $\phi_{2}=0$. When we set $\phi_{2}=\pi/2$, the output intensity at resonant frequency is 0.5 (half-open) and at positive (negative) frequency detuning, the right (left) channel is open. The output switching is completely open when $\phi_{2}=\pi$. The switching contrast ratio at resonance between closed state and open state can be up to 1. Actually, this can be used as the transfer switching from perfect photon absorber to perfect photon transmitter or reflector.

Besides as an optical switching, we notice that the output intensities from two ends of the cavity can always reach the same values under some frequencies (i.e. $\Delta_{p}=0$), while under other frequencies (such as $\Delta_{p}=\pm\Gamma$), output intensities of two ends will be different. This indicates possible enhancement or weakness of intensity correlation or quantum coherence of the two channels. As shown in Fig. \ref{fig-intensity}(a), the output intensities for both channels are minimum thus the intensity correlation between these two channels is at a small value under $\Delta_{p}=0$. With $\phi_{2}=\pi/2$ in figure \ref{fig-intensity}(b), the output intensities of two channels are increased and equal at resonance which indicates intensity correlation is enhanced at larger value for $\Delta_{p}=0$. When $\phi_{2}=\pi$ as in Fig. \ref{fig-intensity}(c), the output intensities are equal and maximum which indicates maximum correlation (robust entanglement) free to the environmental dissipation. This infers that quantum correlation of the two output channels can be created by classic interference between two driving fields as shown in Fig. \ref{fig-intendis}. Since media absorption can be controlled by intracavity phase-dependent EIT and outputs of intracavity mode field can be manipulated by the interference of two coherent input beams induced by driving-field phase, the scheme can be used for manipulating quantum correlation and even the quantum entanglement of the two channels.

\begin{figure}[!hbt]
\centering
\subfigure{
\begin{minipage}{5.9cm}
\includegraphics[width=5cm,bb=100 0 600 510]{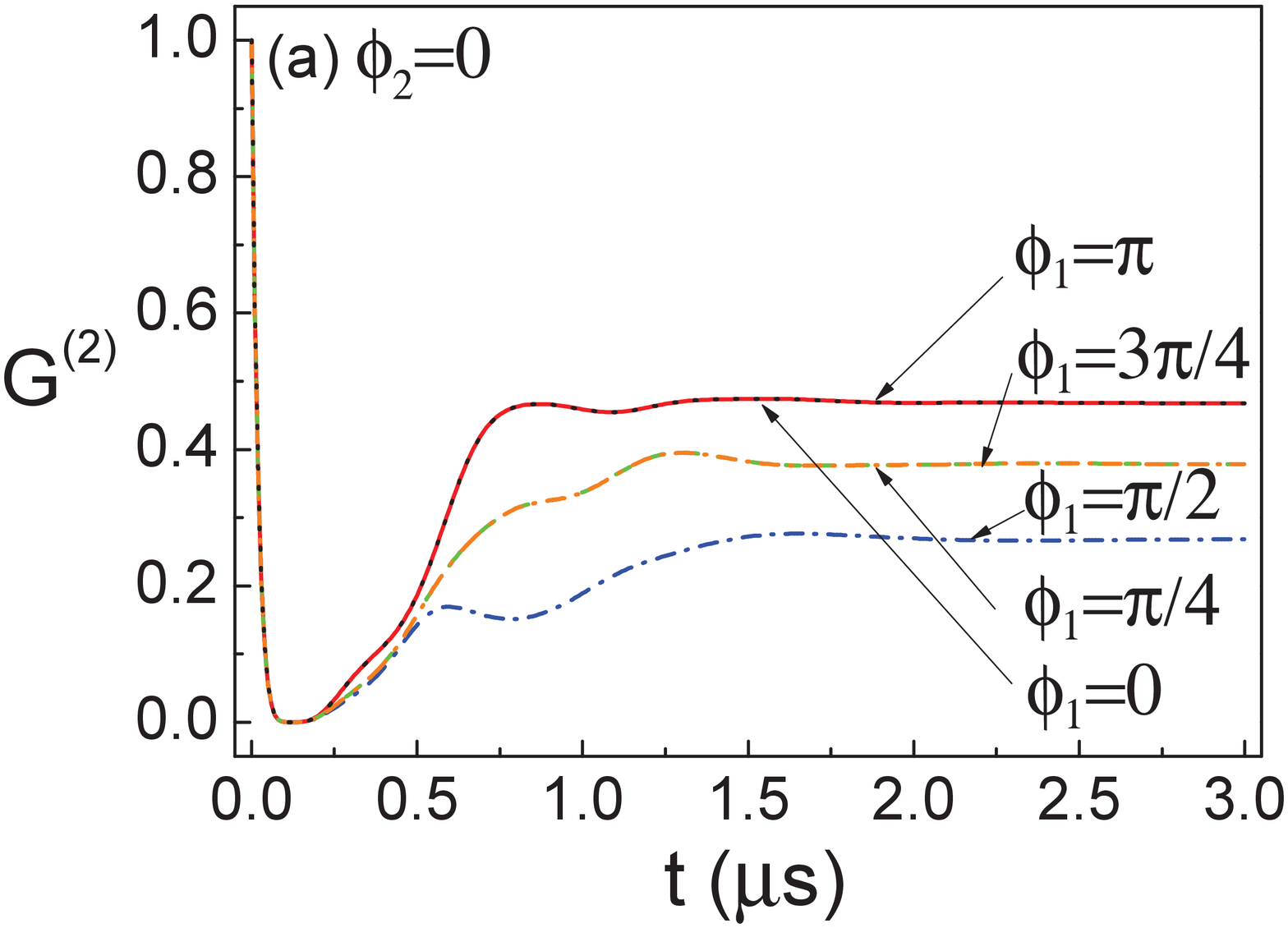}
\end{minipage}}
\subfigure{
\begin{minipage}{5.9cm}
\includegraphics[width=5cm,bb=100 0 600 510]{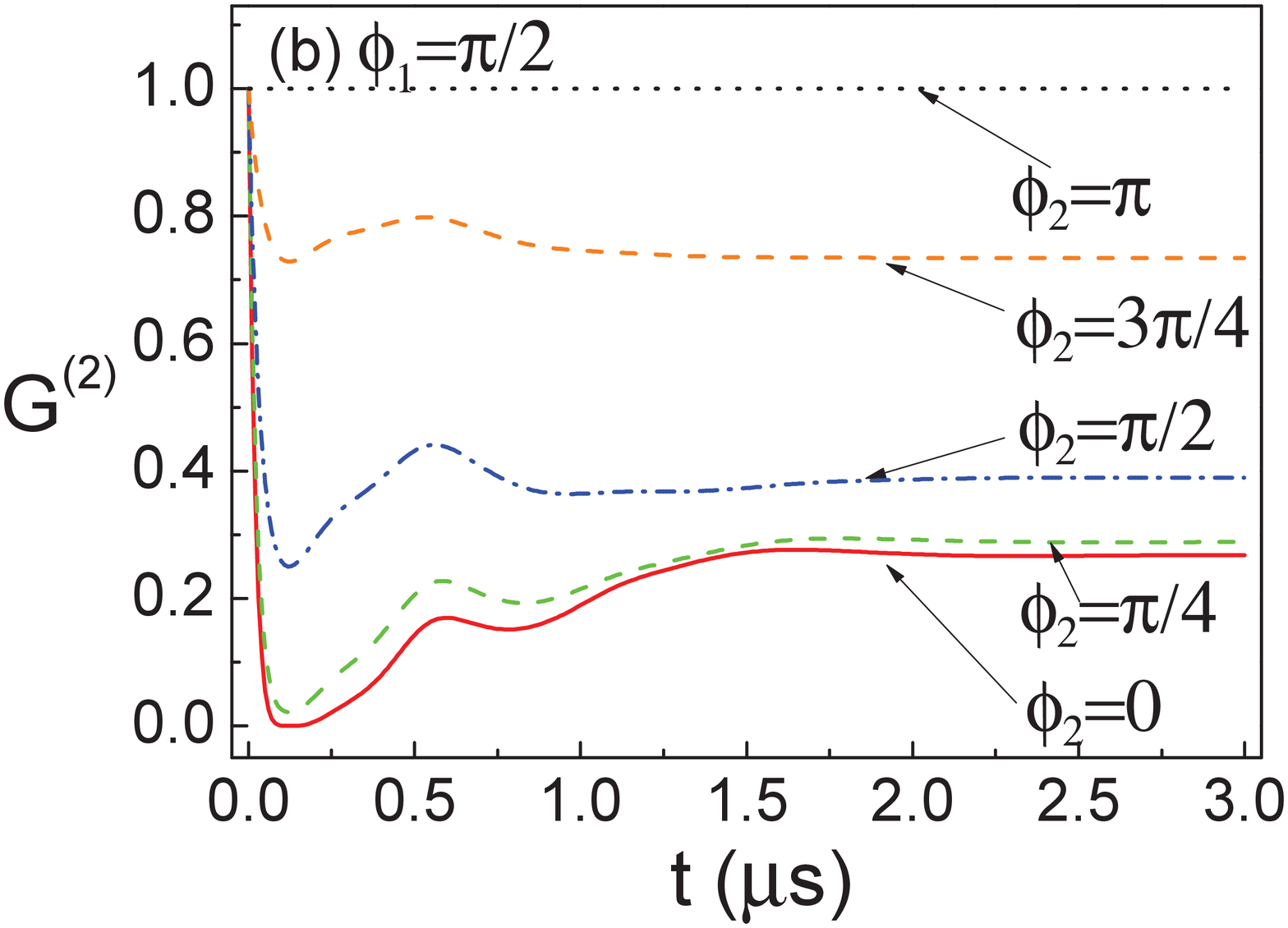}
\end{minipage}}
\caption{Time evolution of $G^{(2)}$ versus (a) $\phi_{1}$ and (b) $\phi_{2}$ at resonant probe frequency. Parameters are (a) $\phi_{2}=0$ and (b) $\phi_{1}=\pi/2$. The other parameters are the same as in Fig. \ref{fig-imx}(b).}\label{fig-3d1evo}
\end{figure}
In order to show the phase-dependent quantum correlation quantitatively, we calculate the second-order correlation \cite{RMP87/1379} between the two channels,
\begin{equation}
G^{(2)}=\frac{\langle a^{\dag}_{out,l}a^{\dag}_{out,r}a_{out,r}a_{out,l}\rangle}{{\langle a^{\dag}_{in,r}a_{in,r}\rangle}^{2}}.\label{G2}
\end{equation}
The time evolution of $G^{(2)}$ at resonant frequency versus $\phi_{1}$ and $\phi_{2}$ are drawn as in Fig. \ref{fig-3d1evo}(a) and \ref{fig-3d1evo}(b), respectively. They are shown obviously that the initial second-order correlation decreases rapidly within $0.12\,\mu s$ (which is in accordance with system loss $\mathcal{L}=\kappa_{l}+\kappa_{r}+\kappa_{atom}=3/2\,\Gamma$, here $\Gamma=6$ MHz for $^{87}$Rb) \cite{RMP87/1379} then starts increasing, and finally reaches a stable value. Since dissipation which causes decoherence of system can be controlled by phase-dependent EIT, the stable value is different for different closed-loop phase $\phi_{1}$ as in figure \ref{fig-3d1evo}(a). $G^{(2)}$ can not get back 1 unless total reflection appears ($\phi_{2}=\pi$) as shown in Fig. \ref{fig-3d1evo}(b). For $\phi_{1}=0$ and $\phi_{1}=\pi$ in Fig. \ref{fig-3d1evo}(a), the stable correlation is around 0.47, while for $\phi_{1}=\pi/2$, the stable value decreases to 0.26, which is accordance with the media absorption property as in Fig. \ref{fig-imx}(b). It reveals that intracavity phase-dependent EIT with driving-field phase can be taken as one source of quantum correlation to conquer decoherence. Similarly in the scheme, quantum correlation can be modulated by standing-wave driving with $\phi_{2}$. As shown in Fig. \ref{fig-3d1evo}(b), although the stable correlation is much lower when $\phi_{1}=\pi/2$ and $\phi_{2}=0$, increasing $\phi_{2}$ from $\phi_{2}=0$ to $\phi_{2}=\pi$, the value of correlation function is increased. The maximum correlation can be obtained when $\phi_{2}=\pi$ where quantum correlation is free to decoherence. These results are consistent with the analysis of output spectra in Fig. \ref{fig-intensity}. It shows that classical interference can also be used to control quantum correlation.

\begin{figure}[!hbt]
\centering
\subfigure{
\begin{minipage}{5.9cm}
\includegraphics[width=5cm,bb=100 0 600 510]{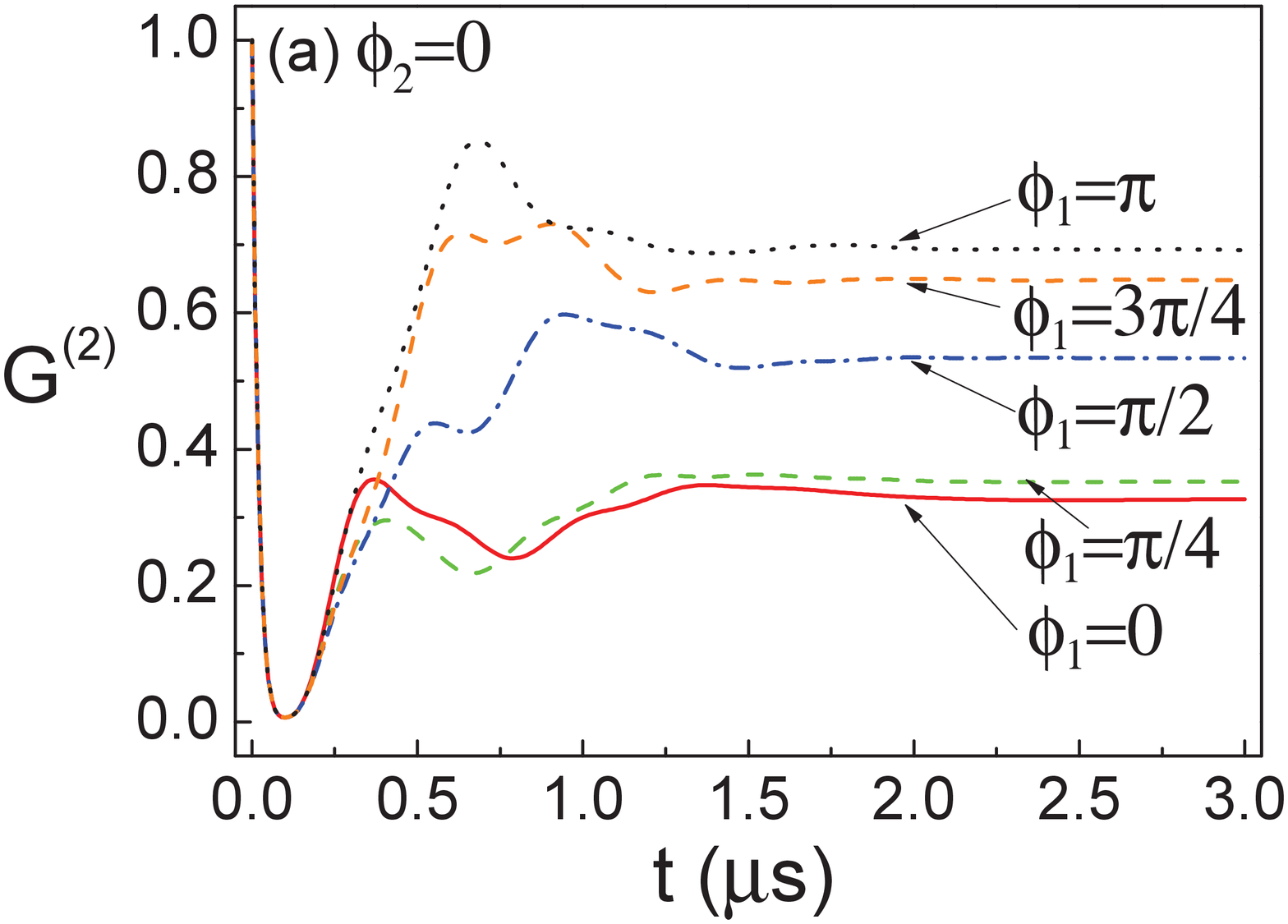}
\end{minipage}}
\subfigure{
\begin{minipage}{5.9cm}
\includegraphics[width=5cm,bb=100 0 600 510]{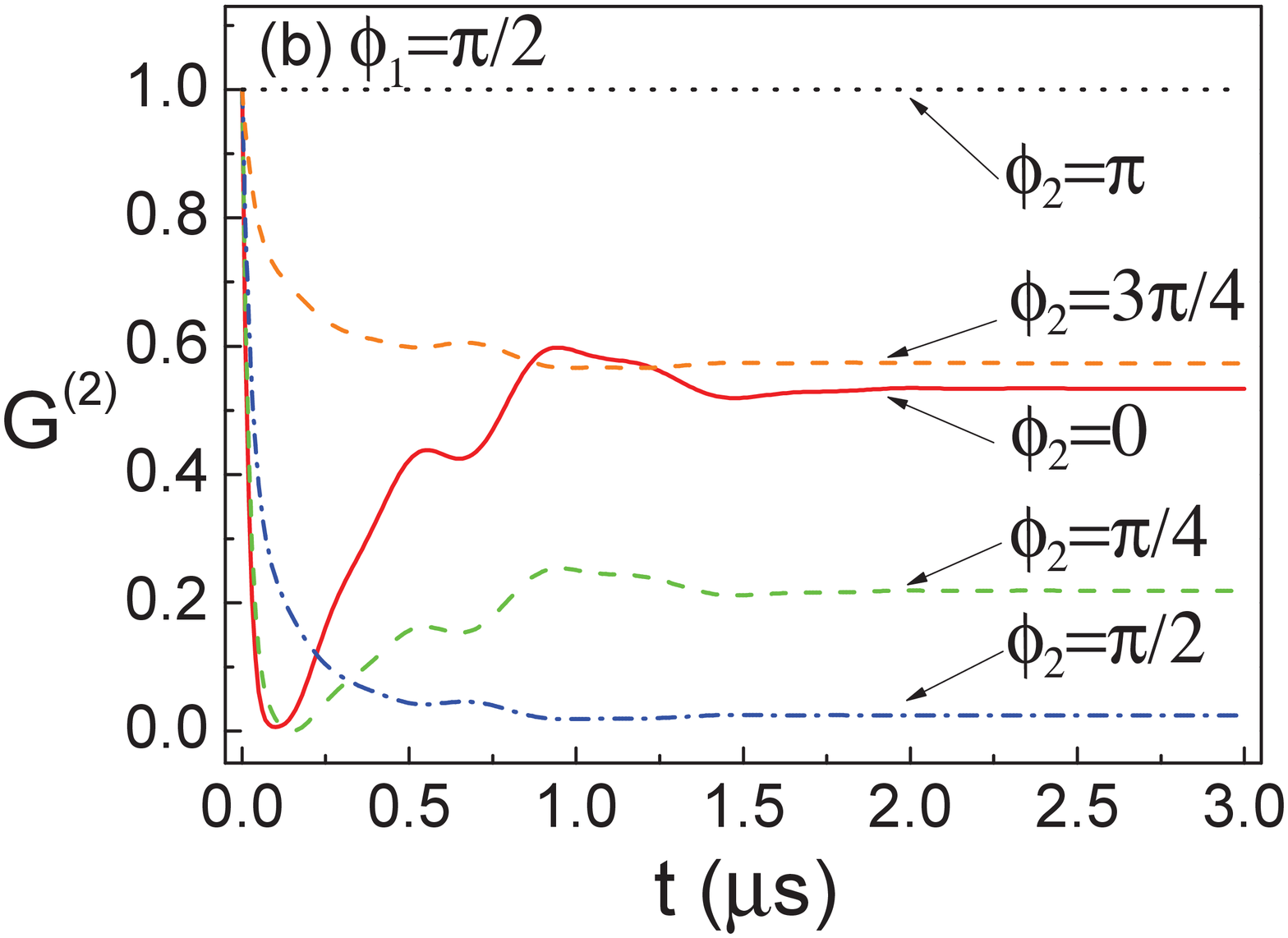}
\end{minipage}}
\caption{Time evolution of $G^{(2)}$ versus (a) $\phi_{1}$ and (b) $\phi_{2}$ at phase-dependent EIT window $\Delta_{p}=\Gamma$. Parameters are (a) $\phi_{2}=0$ and (b) $\phi_{1}=\pi/2$. The other parameters are the same as in Fig. \ref{fig-imx}(b).}\label{fig-3d2evo}
\end{figure}

In the following, we analyze the correlation function at EIT windows ($\Delta_{p}=\pm\Gamma$) with several values of the two phases. Fig. \ref{fig-3d2evo}(a) shows that at $\Delta_{p}=\Gamma$, with increasing $\phi_{1}$ from 0 to $\pi$, stable quantum correlation will be improved from 0.32 to 0.70. Different from that in resonant frequency, the minimum stable correlation is obtained when $\phi_{1}=0$. This is predictable from the analysis of Fig. \ref{fig-imx}(c) and \ref{fig-imx}(d). The output intensities of two channels are equal and enhanced with $\phi_{1}$ being increased from 0 to $\pi$, therefore intensity correlation is enhanced correspondingly. While at $\Delta_{p}=-\Gamma$, the steady correlation is decreasing with changing $\phi_{1}$ from $\phi_{1}=0$ to $\phi_{1}=\pi$ (not shown).

Under $\phi_{1}=\pi/2$, quantum correlation depends on $\phi_{2}$ differently in the situations for $\Delta_{p}=0$ and $\Delta_{p}=\Gamma$ as shown in figures \ref{fig-3d1evo}(b) and \ref{fig-3d2evo}(b). This can be explained according to the analysis of Fig. \ref{fig-intensity}. For $\Delta_{p}=0$, output intensities of two channels under $\phi_{2}=0$ (Fig. \ref{fig-intensity}(a)) are weak but equal and thus there exists minimum enhancement of intensity correlation. The second order correlation can be recovered to a steady value 0.26  under $\Delta_{p}=0$ when $\phi_{1}=\pi/2$, $\phi_{2}=0$ as shown in Fig. \ref{fig-3d1evo}(b). However, for $\Delta_{p}=\Gamma$ in Fig. \ref{fig-intensity}(b) ($\phi_{2}=\pi/2$), the intensities of two output channels are unequal, one of which is strong but the other is weak, thus the correlation can not be recovered (minimum steady correlation close to zero under $\phi_{2}=\pi/2$ at $\Delta_{p}=\Gamma$ as in Fig. \ref{fig-3d2evo}(b)). For $\phi_{2}=0$, $\pi/4$ and $3\pi/4$, the correlation can reach larger steady values and the maximum value for $\phi_{2}=\pi$ can be realized because of the partial or total reflection based on driving-field phases as in Fig. \ref{fig-intendis}.

In a word, in a cavity-atom system, with two-sided coherent driving and closed-loop EIT controlling, phase-dependent quantum correlation can be realized.

\vspace{-5pt}
\section{Conclusions}
In conclusion, we analyze media absorption and wave interference in a four-level atom-cavity system. Based on phase-dependent EIT rather than using large intensity of coupling field, optical switching with high contrast (up to 1) via phase control can be realized. Together with phase-dependent standing wave formed by two coherent driving fields, the total photon absorber or hundred-percent transmitter (reflector) can be obtained.

Due to absorption suppression by modified intracavity EIT and interference enhancement of standing-wave field in the cavity, output intensity can be controlled by closed-loop phase ($\phi_{1}$) and driving-field phase ($\phi_{2}$). Phase-dependent quantum correlation between two output channels can be manipulated. Because of the total reflection from standing-wave field under $\phi_{2}=\pi$, maximum correlation of these two channels can be obtained. By manipulation on quantum correlation, this work provides potential application in realization of controllable entangled photons in cavity system.

\vspace{-5pt}
\section*{Acknowledgment}
We acknowledge support from National Natural Science Foundation of China under Grant No. 11174109.


\end{document}